\pdfoutput=1

\documentclass{article}

\usepackage{microtype}
\usepackage{graphicx}
\usepackage{subfigure}
\usepackage{booktabs} 

\usepackage{hyperref}

\usepackage[accepted]{icml2024}

\usepackage{amsmath}
\usepackage{amssymb}
\usepackage{mathtools}
\usepackage{amsthm}

\usepackage{amsmath,amsfonts,bm}

\def\eqref#1{equation~\ref{#1}}

\def\1{\bm{1}}

\def\rvh{{\mathbf{h}}}

\def\rvz{{\mathbf{z}}}

\def\va{{\bm{a}}}

\def\ve{{\bm{e}}}

\def\vh{{\bm{h}}}

\def\vn{{\bm{n}}}

\def\vv{{\bm{v}}}
\def\vw{{\bm{w}}}
\def\vx{{\bm{x}}}

\def\vz{{\bm{z}}}

\DeclareMathAlphabet{\mathsfit}{\encodingdefault}{\sfdefault}{m}{sl}
\SetMathAlphabet{\mathsfit}{bold}{\encodingdefault}{\sfdefault}{bx}{n}

\def\gA{{\mathcal{A}}}

\def\gE{{\mathcal{E}}}

\def\gI{{\mathcal{I}}}

\def\gK{{\mathcal{K}}}
\def\gL{{\mathcal{L}}}

\def\gS{{\mathcal{S}}}

\def\gV{{\mathcal{V}}}

\def\gX{{\mathcal{X}}}

\def\sR{{\mathbb{R}}}

\def\sZ{{\mathbb{Z}}}

\DeclareMathOperator*{\argmax}{arg\,max}
\DeclareMathOperator*{\argmin}{arg\,min}

\DeclareMathOperator{\sign}{sign}

\usepackage{hyperref}
\usepackage{url}
\usepackage{graphicx}
\usepackage{makecell}
\usepackage{multirow}
\usepackage{enumitem}
\usepackage{bbm}
\usepackage{color,xcolor}
\usepackage{booktabs}
\usepackage{lineno}
\usepackage{array}
\usepackage{ulem}
\usepackage{multicol}
\usepackage{amsmath,amsthm,amsfonts,amssymb,bm}
\usepackage{bbding}
\usepackage{stfloats}
\usepackage[font=small,labelfont=bf]{caption}
\usepackage{tabulary}
\usepackage{tabularx}
\usepackage{float}
\usepackage{colortbl}
\definecolor{Highlight}{rgb}{0.89,0.89,0.94}
\newcommand{\chl}{\cellcolor{Highlight}}
\usepackage[capitalize,noabbrev]{cleveref}

\theoremstyle{plain}

\theoremstyle{definition}

\theoremstyle{remark}

\usepackage[textsize=tiny]{todonotes}

\begin{document}

\twocolumn[
\icmltitle{FoldToken: Learning Protein Language via Vector Quantization and Beyond}



\icmlsetsymbol{equal}{*}

\begin{icmlauthorlist}
\icmlauthor{Zhangyang Gao}{equal,sch}
\icmlauthor{Cheng Tan}{equal,sch}
\icmlauthor{Jue Wang}{sch}
\icmlauthor{Yufei Huang}{sch}
\icmlauthor{Lirong Wu}{sch}
\icmlauthor{Siyuan Li}{sch}
\icmlauthor{Stan Z. Li}{sch}
\end{icmlauthorlist}

\icmlaffiliation{sch}{AI Lab, Research Center for Industries of the Future, Westlake University}

\icmlcorrespondingauthor{Stan Z. Li}{Stan.ZQ.Li@westlake.edu.cn}

\icmlkeywords{Machine Learning, ICML}

\vskip 0.1in
]



\printAffiliationsAndNotice{\icmlEqualContribution} 

\vspace{-5mm}
\begin{abstract}
    Is there a foreign language describing protein sequences and structures simultaneously? Protein structures, represented by continuous 3D points, have long posed a challenge due to the contrasting modeling paradigms of discrete sequences. We introduce \textbf{FoldTokenizer} to represent protein sequence-structure as discrete symbols. This innovative approach involves projecting residue types and structures into a discrete space, guided by a reconstruction loss for information preservation. We refer to the learned discrete symbols as \textbf{FoldToken}, and the sequence of FoldTokens serves as a new protein language, transforming the protein sequence-structure into a unified modality. We apply the created protein language on general backbone inpainting and antibody design tasks, building the first GPT-style model (\textbf{FoldGPT}) for sequence-structure co-generation with promising results. Key to our success is the substantial enhancement of the vector quantization module, Soft Conditional Vector Quantization (\textbf{SoftCVQ}).
\end{abstract}
\vspace{-5mm}

\section{Introduction}
Sequence and structure modeling play a crucial role in protein applications, including mutation prediction \citep{umerenkov2022prostata}, sequence design \citep{ingraham2019generative, jing2020learning, tan2022generative, gao2022alphadesign, zheng2023mmdesign, hsu2022learning, gao2023pifold, gao2023knowledge}, and  function prediction \citep{hu2023learning,zhang2022protein,fan2022continuous, anonymous2024multimodal}. However, the modality gap between sequence and structure usually requires different model architectures, i.e., sequence learners are typically based on transformers, while the structure learners are generally based on graph neural networks (GNNs). Constraints from modality-specific models, especially the structure prior required by GNNs, limit the modeling flexibility and prevent the applications of recent progress, such as GPT \citep{brown2020language}, on protein tasks. Given the success of applying Transformer to ViT \citep{dosovitskiy2020image}, where a image is converted to sequence of patches (image foreign language), we wonder \textit{is there a way to unify the sequence and structure modalities by a protein foreign language?}

Recent work shows promise on structure to sequence transformation. IG-VAE \citep{eguchi2022ig} and FoldingDiff \citep{wu2022protein} suggest converting protein structures as sequences of folding angles and shows great potential in unconditional protein backbone generation. DiffSDS \citep{gao2023diffsds} extends the method in the scenarios of conditional and constrained backbone inpainting. These methods simplify 3D coordinates as angle sequence, allowing advanced transformers to be applied to structural modeling. However, continuous angles resist modeling as discrete symbols, unlike human language, hindering language transformers in auto-regressive generation where regression loss-trained models generally underperform compared to those trained by classification loss. \citep{pintea2023step, stewart2023regression, horiguchi2023streaming}. \textit{Establishing a discrete protein language to bridge protein research with NLP remains an open challenge.}

Inspired by learning discrete vision language \citep{yu2021vector, esser2021taming,lee2022autoregressive}, we create a novel protein foreign language describing sequence and structure simultaneously, dubbled protein language. The key idea is to project the sequence and structure into a joint discrete space via FoldTokenizer then the sequence of discrete codes can serve as a new protein language to unify the sequence and structure modalities. To ensure the created protein language is informationally equivalent to original sequence-structure, we train the FoldTokenizer via reconstruction loss. Once the protein language is learned, it can be used for generative and predictive downstream tasks. Limited to the space and effort, this paper focuses on generative tasks.

Recovering structures from the protein language poses a non-trivial challenge, necessitating the development of new VQ methods. After evaluating vanilla VQ (VVQ) \citep{van2017neural} and recent lookup-free VQ (LFQ) \citep{yu2023language}, we reveal their proficiency in sequence but suboptimal structure reconstruction. We systemically analized their limitations  and strengths, proposing targeted strategies for improvement. \textit{The key to high-quality reconstruction is soft querying across the entire codebook space.}

LFQ \citep{yu2023language} has identified a trade-off between reconstruction and generation tasks - models excelling in reconstruction may underperform in downstream tasks.  We further analize why the phenominon exist and propose imporved VQ method to overcome this challenge. For the first time, the proposed SoftCVQ method achieves good performance on both protein reconstruction and generation tasks, effectively addressing limitations seen in prior approaches.

We assess our model on both reconstruction and generative tasks. Our findings reveal a substantial enhancement in reconstruction quality with the proposed SoftCVQ method surpassing existing VQ methods. Leveraging the learned protein language, we introduce FoldGPT as the pioneering autoregressive sequence-structure co-generation model. Remarkably, FoldGPT outperforms comparable methods relying on sequences of continual angles, providing additional confirmation of the value of discretization. Our evaluation extends to the antibody design ability of FoldGPT, where it surpasses previous methods in the same category.

\section{Related Work}
\paragraph{Protein Sequence-Structure Co-modeling} has emerged as a vibrant area of research, aiming to create joint representations and enable the simultaneous generation of sequences and structures. Co-representation techniques integrate pretrained sequence models with graph neural networks pretrained on protein structures. These fused embeddings find applications in predictive tasks such as binding affinity and stability prediction \citep{zhang2023efficiently, samaga2021scones}. Co-design techniques establish connections between sequence and structure generative models to generate novel sequences and structures from scratch \citep{shi2022protein, song2023functional}. These approaches highlight that sequences and structures contain complementary information, underscoring the need for further advancements in co-modeling methods to address challenges related to both aspects of protein folding and inverse-folding. A notable application of co-modeling is in antibody design \citep{jin2021iterative, luo2022antigen, kong2022conditional, tan2023protein}.

\paragraph{Vector Quantization} is a powerful technique for data compression and generation. Vanilla VQ \cite{van2017neural} compresses latent representations to the nearest codebook vector.  Product quantization (PQ) \citep{5432202, chen2020differentiable, el2022image} decomposes the space into a Cartesian product of low-dimensional subspaces, quantizing each subspace separately for nearest neighbor search. Residual quantization (RQ) \citep{juang1982multiple, martinez2014stacked, lee2022autoregressive, zeghidour2021soundstream} iteratively quantizes vectors and their residuals, representing the vector as a stack of codes. Lookup-free VQ (LFQ) \citep{mentzer2023finite, yu2023language} quantizes each dimension to a small set of fixed values.

\paragraph{Angle-based Structure Representation} Derived from local residue coordinate systems, invariant features such as bond angle, torsion angle, and bond length serve as natural representations of protein structures \citep{eguchi2022ig, wu2022protein, gao2023diffsds, lee2022proteinsgm}. They capture intrinsic geometric properties unaffected by translation or rotation, making them ideal for tasks like inverse folding and antibody design. To fit the paradigm of the transformer, we encode structural information as a sequence of locally folding angles. Formally, we define $f: \vx_i \mapsto \va_i$, transforming Euclidean coordinate $\vx_i$ to invariant representation $\va_i$, i.e., $\va_i = (r_i, \alpha_i, \beta_i)$, where $r_i$, $\alpha_i$ and $\beta_i$ are bond length, bond angle and torsion angle. We illustrate the details of the folding angle computation in the Appendix.\ref{appl:folding_angle}.

\section{Method}

\subsection{Overall framework}
\begin{figure*}[h]
   \centering
   \includegraphics[width=6.5in]{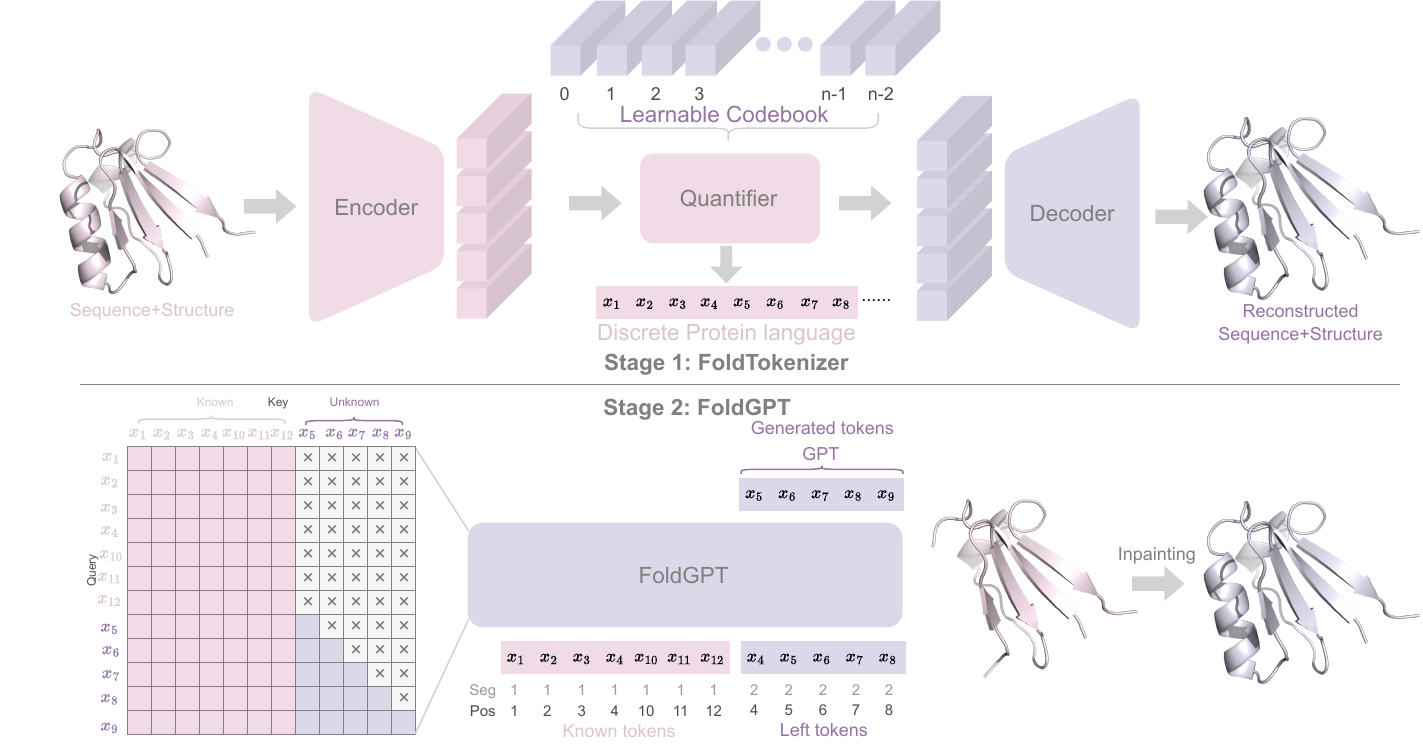}
   \caption{The overall framework of FoldToken. }
   \vspace{-5mm}
   \label{fig:framework}
\end{figure*}

As shown in Figure.\ref{fig:framework}, the overall pipeline include two models: (1) FoldTokenizer, learning the discrete protein language describing sequence and structure (2) FoldGPT, a NLP-style model using protein language and generating protein sequence-structure autoregressively.

\paragraph{FoldTokenizer} includes encoder, quantizer and decoder, to jointly represent the residue types and $C_{\alpha}$ coordinates as discrete latent codes. Given the sequence and structure as $\gS = \{s_i: 1 \leq i \leq n\}$ and $\gA = \{\va_i \in \sR^{3}: 1 \leq i \leq n\}$, with protein length $n$. We formulate FoldTokenizer as:

\begin{equation}
   \label{eq:FoldTokenizer}
   \begin{cases}
      \rvh \sim  q_{enc}(\rvh| \gS, \gA)\\
      \rvz = Q(\rvh)\\
      \hat{\gS}, \hat{\gA} = p_{dec}(\gS, \gA|Q^{-1}(\rvz))\\
   \end{cases}
\end{equation}

The encoder $q_{enc}(\cdot)$ is a transformer-based model that maps protein sequence and structure into a continuous latent embedding $\rvh = \{\vh_1, \vh_2, \cdots, \vh_n\} \in \sR^{n,d}$. The quantizer $Q: \vh \mapsto \vz$ converts continual embeddings $\vh$ as discrete codes $\vz \in \sZ^{n,1}$, whose reverse operation is $Q^{-1}: \vz \mapsto \hat{\vh}$. The decoding transformer $p_{dec}(\cdot)$ reconstructs the sequence and structure from the reversed latent embedding $\hat{\vh}$.


\paragraph{FoldGPT} We apply the created protein language to the problem of protein inpainting. Denote the masked backbone atoms as $\mathcal{M} = \{ (\boldsymbol{a}_i, s_i) \}_{i \in \gI_m}$, the known backbone atoms as $\mathcal{K} = \{ (\boldsymbol{a}_i, s_i) \}_{i \in \gI_k}$, where $\boldsymbol{a}_i$ and $s_i$ are the coordinates and residue type of the $i$-th $C_{\alpha}$ atom. Here, $\gI_m$ and $\gI_k$ denote the indices of unknown and known residues, respectively. The objective of protein inpainting is to generate $\hat{\mathcal{M}}$ using a learnable function $f_{\theta}$ conditioned on $\mathcal{K}$:
\begin{equation}
   f_{\theta}: \gK \rightarrow \mathcal{M}
\end{equation}
Using FoldTokenizer, we encode structure-sequence $\{ (\va_1, s_1), (\va_2, s_2), \cdots, (\va_n, s_n) \}$ as discrete codes $\{ z_1, z_2, \cdots, z_n \}$.  Subsequently, we mask out and regenerate $\{ z_i \}_{i \in \gI_m}$  conditioned on $\{ z_i \}_{i \in \gI_k}$. A distinctive feature compared to previous work is our transformation of the multimodal sequence-structure into a discrete foreign language. This allows for a pure transformer model in structure-sequence co-design, eliminating the need for SE-(3) models and diffusion strategies.  As shown in Fig.\ref{fig:framework},  FoldGPT draws inspiration from GLM and utilizes full attention over all known contexts before generating masked regions in an autoregressive manner. To differentiate between the known and unknown segments, we utilize segment encoding in conjunction with position encoding.  We formulate FoldGPT as
\begin{equation}
   \max_{\theta} \mathbb{E} \sum_{k \in \gI_m}{\log p_{\theta} (z_k | \{z_i\}_{i \in \gI_k}, \{z_j\}_{j<k, j \in \gI_m}) }
\end{equation}

\subsection{FoldTokenizer: Encoder \& Decoder}
\paragraph{Protein Encoder} The encoder transformer uses rotary position encoding and projects the protein sequence $\gS$ and structure $\gX$ as continuous latent embedding $\rvh \in \sR^{d}$. We fuse the sequence-structure embeddings by simply addition:
\begin{equation}
   \label{eq:embedding}
   \begin{cases}
      \rvh^s = \text{Embedding}(\gS) \\
      \rvh^x = \text{Linear}(\gX) \\
      \rvh = \text{Transformer}_{enc}(\rvh^s + \rvh^x) 
   \end{cases}
\end{equation}

\paragraph{Protein Decoder} The decoder has the same architecture as encoder and reconstructs sequence-structure from the discrete latent codes $\rvz$ by separate predictive heads:
\begin{equation}
   \label{eq:decoder}
   \begin{cases}
      \rvh^{dec} = \text{Transformer}_{dec}(Q^{-1}Q(\rvh))\\ 
      \hat{\gS} = \text{Linear}_{s}(\rvh^{dec}) \in \sR^{n,20} \\
      \hat{\gX} = \text{Linear}_{x}(\rvh^{dec}) \in \sR^{n,1} \\
   \end{cases}
\end{equation}

\paragraph{Reconstruction Loss} The learning objective is to recover protein sequence-structure like VQ-VAE. We use cross-entropy loss for residue type classification:
\begin{align}
    \label{eq:reconstruction_seq}
    \gL_{s} &= \log p(\gS | \rvz) 
         =  -\sum_{i=1}^{n} s_i \log(\hat{s}_i) 
\end{align}
To handle the periodicity of the angles ($\alpha_i, \beta_i$), we minimize their trigonometric values, i.e., $l_{\alpha_i} = || \sin{\alpha_i} - \sin{\hat{\alpha}_i} ||_2^2 +|| \cos{\alpha_i} - \cos{\hat{\alpha}_i} ||_2^2$ for $\alpha_i$, the structure loss is
\begin{align}
   \label{eq:reconstruction_struct}
   \gL_{a} = \log p(\gX | \rvz) = \sum_{i=1}^{n}\left(|| r_i - \hat{r}_i ||_2^2 + l_{\alpha_i} + l_{\beta_i} \right)
\end{align}
The overall reconstruction is a weighted version:
\begin{align}
    \label{eq:reconstruction_loss}
    \gL_{rec} = \lambda_{s} \gL_{s} + \lambda_{a} \gL_{a} 
 \end{align}

 \subsection{Challenges on Vector Quantization}
\begin{figure}[h]
     \centering
     \includegraphics[width=3.0in]{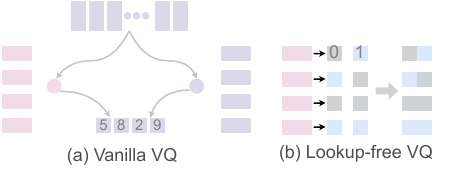}
     \vspace{-3mm}
     \caption{Baseline vector quantization methods. }
     \vspace{-5mm}
     \label{fig:previous_vq}
 \end{figure}

In this section, we thoroughly examine existing vanilla quantizers and lookup-free quantizers, as shown in Fig.\ref{fig:previous_vq}. By understanding these limitations and strengths, we aim to pave the way for further improvements in the field.
 
 \paragraph{Vector Quantization Problem} The quantizer $Q: \vh \mapsto z$ converts continuous embedding $\vh$ as a discrete latent code $z$, named VQ-ID. The de-quantizer $Q^{-1}: z \mapsto \hat{\vh}$ recover continual embedding $\hat{\vh}$ from $z$. The problem formulation is
 \begin{equation}
    \begin{cases}
       z &= Q(\vh)\\
       \hat{\vh} &= Q^{-1}(z)
    \end{cases}
 \end{equation}
 \vspace{-5mm}
 \paragraph{Strategy A: Vanilla Quantizer (VVQ)} Given input $\vh$, the vanilla quantizer finds the nearest codebook vector $\vv$ from codebook $\gE = \{ \vv_k \}_{k=1}^m$. The index of $\vv$ serves as the discrete latent code $z$, dubbled VQ-ID of $\vh$.
 \begin{equation}
    \begin{cases}
       z_i &= Q(\vh_i) = \argmin_{j} ||\vh_i - \vv_j||_2 \\
       \hat{\vh}_i &= Q^{-1}(z_i) = \vv_{z_i} 
    \end{cases}
    \label{eq:quantizer}
 \end{equation}
 To improve the learning stability of VVQ, we utilize the exponential moving average (EMA) mechanism for codebook updating and normalize the continuous embeddings $\vh$ and $\vv$ to the unit hypersphere before quantization: $\vh \leftarrow \vh / ||\vh||_2$, $\vv \leftarrow \vv / ||\vv||_2$. In addition to the gradient update, the EMA moves $\vv$ to the center of its nearest top-$k$ embeddings $\{\vh_j: 1 \leq j \leq k\}$, when $\vv_i$ deviates from the data distribution. This process is described by $\vv_{i}  \leftarrow \tau \vv_{i} +  (1-\tau)(\vh_i - \vv_{i})$. $\tau = 0.95 \quad \text{if} \quad || \frac{1}{k}\sum_{j=1}^k \vh_j - \vv_{i}||_2 > \epsilon$, otherwise $\tau =0.0$. Note that $\tau, k$, and $\epsilon$ are the moving average weight, the number of nearest embeddings, and the threshold for out-of-distribution embeddings, respectively. We set $k=5$ and $\epsilon=0.1$. We apply the VQ loss for aligning $\vh$ and $\vv$, denoted as $\mathcal{L}_{vq}= ||\text{sg}(\vh) - \vv||_2^2 + 0.25 ||\vh - \text{sg}(\vv)||_2^2 $. 
 

 \paragraph{\underline{Limitations of VVQ}} The vanilla quantizer directly copies the gradient of $\hat{\vh}_i$ to $\vh_i$, i.e., $\frac{\partial L}{\partial \vh_i} \leftarrow \frac{\partial L}{\partial \hat{\vh}_i}$, resulting in the \textbf{gradient mismatching} between embedding $\frac{\partial L}{\partial \vh_i}$ and $\vh_i$. In addition, the higher reconstruction quality does not necessarily lead to better generation performance on downstream tasks. For example, \citep{yu2023language} points out that reconstruction and generation may be contradicted: \textit{Enlarging the vocabulary can improve reconstruction quality. However, such improvement only extends to generation when the vocabulary size is small, and a very large vocabulary can actually hurt the performance of the generative model.} Why does the contradiction occur? We attribute the intrinsic reason to \textbf{semantic irrelevance} between continuous and discrete representations and the \textbf{large class space}. We summarize the limitations as follows:
 \begin{enumerate}[itemsep=1pt]
    \item \textbf{Gradient Mismatching} As the VQ-IDs $\{z_i\}_{i=1}^m$ are non-differentiable, the vanilla quantizer directly copies the gradient of $\hat{\vh}_i$ to $\vh_i$, i.e., $\frac{\partial L}{\partial \vh_i} = \frac{\partial L}{\partial \hat{\vh}_i}$. However, $\hat{\vh}_i$ generally does not equal $\vh_i$, resulting in the mismatching between embedding $\vh_i$ and its gradient $\frac{\partial L}{\partial \vh_i}$.
    \item \textbf{Semantic Irrelevance} The discrete and continuous representations exhibit no correlation, implying the absence of any association between $(\vh_i-\vh_j)$ and $(z_i - z_j)$. The irrelevance means that the generative model should accurately predict the exact $z_i$ for recovering $\vh_i$, despite the high similarity between $z_i$ and $z_j$. This further increase the differenty of generation.
    \item \textbf{Large Class Space} Regarding the generative model, each VQ-ID $z_i$ represents a class index. The large codebook size of $m$ can indeed pose challenges in generating the "protein language".
 \end{enumerate}
 
 \paragraph{Strategy B: Lookup-Free Vector Quantizer (LFQ)} This method \citep{yu2023language} address the semantic mismatching and large class space problems. Denote $\ve_j \in \sR^{ d}$ as the one-hot vector whose $j$-th element is 1 and $d = \log_2 m$. The key idea is to project the continuous $\vh_i$ into the space of $ \mathcal{V} = \{ \sum_{j=1}^{d} \left[ w_{ij} \ve_j + (1-w_{ij}) (-\ve_j) \right]: w_{ij} \in \{0, 1\} \} $:
 \begin{equation}
    \begin{cases}
       \hat{\vh}_i = \sign{(\vh_i)} = \sum_{j=1}^{d} \left[ u(h_{ij}) \ve_j + (1-u(h_{ij})) (-\ve_j) \right]  \\
       \text{Bit}^{-1}(\vz, d) := \sum_{j=1}^{d} 2^{j-1} z_j\\
       \text{Bit}(z, d) := [\lfloor z/2^{d-1} \rfloor, \lfloor z/2^{d-2} \rfloor, \cdots, \lfloor z/2^{0} \rfloor]^T\\
       z_i = \text{Bit}^{-1} (u{(\hat{h}_{i,j})})\\
    \end{cases}
    \label{eq:LFQ}
 \end{equation}
 The LFQ transforms latent embedding $\vh_i$ into sign vector $\hat{\vh}_{i} \in \{-1,1\}^d$.  $\sign(x) = 1 \quad \text{if} \quad x \geq 0$ and $\sign(x) = -1 \quad \text{if} \quad x < 0$. $u(x) = (\sign(x)+1)/2$. We refer to $u(\hat{\vh}_{i})$ as the binary VQ-ID vector, and its decimal form is used as the VQ-ID $z_i$. The functions $\text{Bit}^{-1}(\cdot, \cdot)$ and $\text{Bit}(\cdot, \cdot)$ define the transformation between the binary VQ-ID vector and the VQ-ID.
 
 \paragraph{Strength of LFQ} The binary VQ-ID $u(\hat{\vh}_{i})$ directly captures the sign information of $\vh_i$, establishing a semantic relevance between the discrete and continuous representations. This semantic relevance offers several advantages for downstream tasks, including:
 \begin{enumerate}
    \item \textbf{Robust Generation}: Accurate prediction of all bits is unnecessary for the model. Majority of correct "key" bits ensures similar overall semantics, leading to reasonable reconstruction of proteins.
    \item \textbf{Simplified Classification}: Unlike VVQ that predicts tokens from a space of size $m$, the LFQ allows us to decompose the problem as $\log_2{m}$ binary classifications. 
 \end{enumerate}
 LFQ addresses the issue of semantic irrelevance and make the generation more robust and easy to be optimized.

 \paragraph{\underline{Limitations of LFQ}} LFQ fails to address the issue of gradient mismatching and and introduces a new challenge of information bottleneck.Typically, the codebook space is chosen from options such as $2^8$ (int8), $2^{16}$ (int16), $2^{32}$(int32), and $2^{64}$(int64) due to accommodate the required bits for storing a VQ-ID. However, for a fair comparison to VVQ and effective data compression, only $2^8$ and $2^{16}$ are considered, resulting in a hidden space size of 8 or 16 in lookup-free VQ. This low dimensionality poses the problem of information bottleneck, which hampers accurate reconstruction.

 \subsection{How to improve VQ methods?}
 To address the aforementioned problems, we introduce three VQ methods, e.g., soft quantizer (SoftVQ), soft group quantizer (SoftGVQ) and soft conditional quantizer (SoftCVQ). In Table.~\ref{tab:quantizer}, we summarize addressed issues of VQ methods, i.e., \textbf{g}radient \textbf{m}ismatching (GM), \textbf{s}emantic \textbf{i}rrelevance (SI), \textbf{m}ulticlass \textbf{c}lassification (MC), and \textbf{i}nformation \textbf{b}ottleneck (IB), and the strength of \textbf{s}oft \textbf{g}lobal \textbf{q}uerying (SGQ). Readers will understand SGQ from strategy E.
 
 \begin{table}[h]
    \caption{Comparison of different vector quantization methods.}
    \label{tab:quantizer}
    \centering
    \resizebox{1.0 \columnwidth}{!}{\begin{tabular}{lccccc}
       \toprule
    VQ Method               & w/o GM & w/o SI & w/o MC & w/o IB & w SGQ\\ \midrule
    Vanilla     & &   &   &  \checkmark \\
    Lookup-free & &   \checkmark &  \checkmark \\ \midrule
    \chl SoftVQ (our)        &\chl \checkmark & \chl  & \chl & \chl \checkmark & \chl \checkmark\\
     \chl SoftGVQ (our) & \chl \checkmark & \chl \checkmark & \chl \checkmark & \chl \checkmark & \chl\\ 
    \chl SoftCVQ (our) & \chl \checkmark & \chl \checkmark & \chl \checkmark & \chl \checkmark & \chl \checkmark\\ \bottomrule           
    \end{tabular}}
 \end{table}

 \begin{figure}[h]
   \centering
   \includegraphics[width=3.3in]{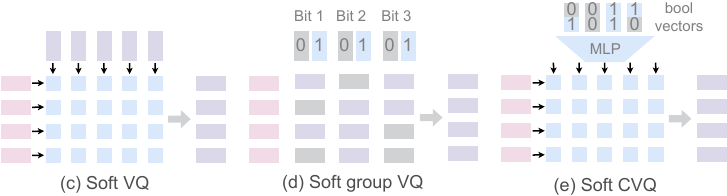}
   \vspace{-3mm}
   \caption{Proposed vector quantization methods. }
   \label{fig:proposed_vq}
\end{figure}

 \paragraph{Strategy C: Soft Vector Quantizer (SoftVQ vs VVQ)} Instead of selecting the closest codebook vector, the soft quantizer utilizes a differentiable attention operation to query the entire codebook $\gE= \{ \vv_k \}_{k=1}^m$:
\begin{equation}
   \begin{cases}
      a_{ij} &= \frac{\exp{(\vh_i^T \vv_j/T)}}{\sum_{j=1}^m \exp{(\vh_i^T \vv_j/T)}}\\
      \hat{\vh}_i &= \sum_{j=1}^m a_{ij} \vv_j\\
      z_i &= \argmax_{j} a_{ij}\\
   \end{cases}
\end{equation}
The temperature parameter $T$ plays a crucial role in controlling the smoothness of the attention map. When $T \rightarrow 0$, the attention map becomes a one-hot vector, which is equivalent to the VVQ. During the training process, as $T$ gradually decreases from 1.0 to $1e^{-5}$, the model automatically learns a sharp attention map for querying the "nearest" codebook vector. The differentiable attention operation avoids gradient mismatch problem. During the inference phase, we employ random sampling to obtain the discrete latent codes $z_i$ from the distribution $z_i \sim \text{Multinomial}([a_{i,0}, a_{i,1}, \cdots, a_{i,M}])$.

\paragraph{Strategy D: Soft Group Vector Quantizer (SoftGVQ vs LFQ)} Similar to LFQ, the soft group quantizer projects the continuous embedding $\vh_i$ into the space of $ \gV =  \{ \sum_{j=1}^{\log_2(m)} \left[ w_{ij} \vv_j + (1-w_{ij}) \dot{\vv}_j \right]: w_{ij} \in [0, 1] \} $, where $m$ is the codebook size, $\vv_j$ and $\dot{\vv}_j$ are learnable basis vectors in the $j$-th group. The $w_{ij}$ is computed by soft attention, i.e., $w_{ij} = \frac{\exp{(\vh_i^T \vv_j/T)}}{\exp{(\vh_i^T \vv_j/T)}+\exp{(\vh_i^T \dot{\vv}_j/T)}}$. During training, we gradually reduce $T$ from 1 to $1e-5$. Define $\vw = [w_{i,1}, w_{i,2}, \cdots, w_{i,\log_2(m)}]$, the binary VQ-ID vector is $u(\vw-0.5)$. The recovered embedding $\hat{\vh}_i$ and VQ-ID $z_i$ could be computed by
\begin{equation}
   \begin{cases}
      \hat{\vh}_i = \sum_{j=1}^{\log_2(m)} [w_{ij} \vv_j + (1-w_{ij}) \dot{\vv}_j] \\
      z_i = \text{Bit}^{-1}(u(\vw-0.5))\\
   \end{cases} 
\end{equation}

The soft group quantizer offers several advantages: 
\begin{enumerate}
   \item The basis vectors ($\vv_j$ and $\dot{\vv}_j$) are learnable and data-dependent, and do not limited to binary one-hot basises ($\ve_j$ and $-\ve_j$) used by LFQ. This allow the model to capture the principle directions of the data distribution.
   \item Within each group, a soft quantization strategy is employed to select the nearest codebook vector, addressing the issue of gradient mismatching. 
   \item The hidden space dimension equal to the that of basis vectors, which is not constrained to be $\log_2(m)$ as in LFQ, elleviating the issue of information bottleneck.
\end{enumerate}

\paragraph{Strategy E: Soft Conditional Quantizer (SoftCVQ, Final Version)} SoftVQ and SoftGVQ present improvements over VVQ and LFQ, respectively. However, we have observed a trade-off between high-quality reconstruction (SoftVQ) and generation (SoftGVQ). While the binary VQ-IDs of SoftGVQ are beneficial for generation tasks, they perform relatively poorer in terms of reconstruction compared to SoftVQ. To strike a balance between reconstruction and generation capabilities, we propose a novel approach called the Soft Conditional Vector Quantizer (SoftCVQ). The insights comes from following obervations:
\begin{enumerate}
   \item The key to achieving high-quality reconstruction lies in employing soft attention to query the entire codebook space, i.e., soft global querying $\rightarrow$ inspired by SoftVQ.
   \item The binary form of VQ-ID facilities generation $\rightarrow$ inspired by SoftGVQ and LFQ.
\end{enumerate}

The proposed SoftCVQ allow attention query over the whole codebook while keeping binary VQ-IDs. A straightforward idea is to use a codebook comprising fixed binary vectors, i.e., $\gV = \{\text{Bit}(z, \log_2(m)) \in \sR^{\log_2(m)}\}_{z=1}^m$, and then applies soft attention between latent embedding $\vh$ and all the codebook vectors $\gV$. However, this approach suffers from information bottleneck due to the small latent dimension $\log_2(m)$. To address this challenge, we introduce a conditional network that projects $\log_2(m)$-dimensional binary vectors of $\gV$ into $d$-dimensional vectors:
\begin{equation}
   \vv_j = \text{ConditionNet}(\text{Bit}(z, \log_2(m)) ) 
\end{equation}
the $\text{ConditionNet}: \sR^{\log_2(m)} \rightarrow \sR^{d}$ is a MLP. If the codebook size is $2^{16}$, the MLP projects 65536 16-dimension boolvectors into 65536 d-dimension vectors. Then we apply soft vector quantization between latent embedding $\vh_i$ and $\{ \vv_z \}_{z=1}^m$ to obtain the quantized embedding $\hat{\vh}_i$
\begin{equation}
   \hat{\vh}_i = \text{SoftVQ}(\vh_i, \{ \vv_z \}_{z=1}^m) 
\end{equation}
During evaluation, when the nearest codebook vector to $\hat{\vh}_i$ is $\vv_z$, the corresponding VQ-ID is exactly $z$, where its binary VQ-ID vector is $\text{Bit}(i, \log_2(m))$.

\section{Experiments}
We conduct experiments to answer the following questions:
\begin{itemize}[leftmargin=5.5mm]
   \item \textbf{Reconstruction (Q1):} Do the proposed methods outperform baselines on  reconstruction quality?
   \item \textbf{Backbone Inpainting (Q2):} How does the learned protein language perform in the generation task?
   \item \textbf{Application (Q3):} How about applying protein language to antibody design?
   \item \textbf{Ablation (Q4):} What are the key factors contributing to the effectiveness of SoftCVQ? Refer to the Appendix.\ref{appl:ablation}.
\end{itemize}

\subsection{Datasets}
\paragraph{cAF2DB for VQ Pretraining} We employ cAF2DB, a clustered version of the AlphaFold Uniprot v3 database, for VQ pretraining. The original AF2DB contains a large number of structures (214,684,311), which presents computational challenges. To overcome this, we utilize cAF2DB \citep{barrio2023clustering} consisting of 2.27M structural clusters. The representative protein structure with the highest pLDDT score is selected from each cluster, while structures with lengths below 30 or pLDDT scores lower than 70 are excluded. This selection process yields a final set of 1,323,729 proteins for VQ pretraining.

\paragraph{CATH4.3 for Backbone Inpainting} For the task of backbone inpainting, we employ CATH4.3. To construct the train, validation, and test sets, we utilize the CAT code to randomly partition proteins in a ratio of 95:2:3. This results in a train set with 30,290 samples, a validation set with 638 samples, and a test set with 957 samples. The test set comprising 957 samples is used for evaluating reconstruction performance. As to the evaluation of backbone inpainting, we apply a filtering criterion based on proteins that can be folded by ESMFold. Specifically, we choose proteins with an average pLDDT score of 0.9 or higher to obtain the inpainting test set comprising 117 proteins.

\paragraph{SAbDab for Antibody Design} We fine-tune FoldGPT on the Structural Antibody Database (SAbDab) \citep{dunbar2014sabdab} and assess its efficacy in antibody design. Following the protocols of \citep{jin2021iterative, kong2022conditional, tan2023protein}, we cluster data by CDRs, resulting in 765 clusters for CDR-H1, 1093 clusters for CDR-H2, and 1659 clusters for CDR-H3. Each cluster is then subdivided into training, validation, and testing sets in an 8:1:1 ratio. Our reported results are based on a 10-fold cross-validation.

\begin{table*}[t]
   \centering
   \resizebox{2 \columnwidth}{!}{
   \begin{tabular}{ccccccccccccccccccc}
   \toprule
   Method  & \multicolumn{2}{c}{Global Metrics} & \multicolumn{2}{c}{Success Rates}  & \multicolumn{6}{c}{Local Metrics}  \\ 
       & Rec    & TMScore & Rec $\geq 0.95$   & TMScore  $\geq 0.5$  & $L_r$  & $L_{\alpha}$ & $L_{\beta}$  & $\max L_r$  & $\max L_{\alpha}$ & $\max L_{\beta}$   \\ \midrule
   Vanilla    & \underline{0.9757} & 0.4224 & \underline{0.9791} & 0.1693 & 0.0966 & 0.0602 & 0.0783 & 2.1790 & 0.5605 & 0.3723 \\
   LFQ    & 0.9482 & 0.4385 & 0.7534 & 0.2894 & 0.1066 & 0.0701 & 0.1079 & 2.2068 & 0.5856 & 0.5257 \\
   \chl SoftVQ (our)  & \chl 0.9713 &\chl \textbf{0.7616} &\chl 0.9781 &\chl \textbf{0.9530} &\chl \underline{0.0683} &\chl \textbf{0.0438} &\chl \textbf{0
   0232} &\chl \underline{1.5647} &\chl \textbf{0
   4989} &\chl \textbf{0.1439} \\
   \chl SoftGVQ (our)   &\chl 0.9743 &\chl 0.5703 &\chl \textbf{0.9875} &\chl 0.5120 &\chl 0.0936 &\chl 0.0616 &\chl 0.0619 &\chl 2.0090 &\chl 0.5655 &\chl 0.3054 \\ 
   \chl SoftCVQ (our) &\chl \textbf{0.9880} &\chl \underline{0.7470} &\chl 0.9603 &\chl \underline{0.9498} &\chl \textbf{0.0523} &\chl \underline{0.0447} &\chl \underline{0.0284} &\chl \textbf{0.8274} &\chl \underline{0.5334} &\chl \underline{0.1646}\\
   \bottomrule           
   \end{tabular}}
   \caption{Performance of protein VQ models on CATH4.3 test set. We highlight the \textbf{best} results and \underline{suboptimal}.}
   \label{tab:reconstruction}
   \vspace{-5mm}
\end{table*}

\subsection{Reconstruction (Q1)}
\textit{Do the proposed methods outperform baselines on  reconstruction quality?}

\paragraph{Setting} We conduct fair comparison for different VQ models by using the same codebook space of $2^{16}$, code vector dimension of 32 and the same encoder-decoder architectures. Both the encoder and decoder adopt ESM-35M bert transformer.  The FoldTokenizer is pretrained on cAF2DB for transforming protein sequence-structure into sequence of VQ-IDs. The evaluation is conducted on real-word proteins, i.e., the whole test set of CATH4.3 with size of 957 samples. With BF16 precision training in DeepSpeed, the model is trained for 15 epochs using the OneCycle scheduler and AdamW optimizer. The batch size is set to 128, the learning rate is 0.0001, and the padding length is 512.

\paragraph{Baselines} Currently, no method exists for protein sequence-structure quantization and reconstruction. We conduct a comparative analysis of five VQ methods, namely VVQ \citep{van2017neural}, LFQ \citep{mentzer2023finite, yu2023language}, SoftVQ, SoftGVQ, and SoftCVQ, all implemented under the same encoder-decoder architecture. It's noteworthy that VVQ is the most widely used method, whereas LFQ stands as the recent state-of-the-art.

\paragraph{Metrics} In evaluating sequence reconstruction, we quantify the recovery rate. The assessment of recovered structure quality involves both global structure similarity (TMScore) and local residue reconstruction (recovering loss). For a set of $b$ proteins with lengths ${ n_1, n_2, n_b}$, we present the average recovering loss per residue ($L_{r}, L_{\alpha}, L_{\beta}$) and maximum recovering loss per residue ($\max L_{r}, \max L_{\alpha}, \max L_{\beta}$), with detailed definitions available in the appendix. To aid bioscientists in algorithm selection, we introduce the reconstruction success rate. A reconstruction is deemed successful if the reconstructed data achieves a minimum of 95\% recovery and 50\% TMScore compared to the reference.

\textit{We present the reconstruction results in Table.\ref{tab:reconstruction} and have following findings:} 

\textbf{Reconstructing structure poses a non-trivial challenge.} While all VQ methods demonstrate satisfactory performance in sequence recovery, the vanilla VQ and LFQ models encounter difficulties in accurately reconstructing protein structure. Specifically, these models exhibit an average TMScore below 0.5 and a structure success rate of less than 30\%.

\textbf{The key to high-quality reconstruction is soft querying across the entire codebook space.} SoftVQ, SoftCVQ, and SoftGVQ achieve TMScores above 0.5 and structure success rates exceeding 50\%. It's worth noting that SoftVQ and SoftCVQ exhibit superior reconstruction performance than others, with success rates around 95\%. Given this, we consider them to be reliable protein quantization tools.

\textbf{VQ methods proves highly effective in data compressing.} When considering the storage of the $C_{\alpha}$ atom, the entire CATH4.3 dataset (2.7GB pdb files) could be compressed to discrete protein language sentences (18MB) using SoftCVQ.

In Fig.\ref{fig:VQ_example} in Appendix \ref{appl:visualization}, we present visual examples of reconstructed proteins, accompanied by TMScore and recovery rate notations. The proposed methods, including SoftGVQ, SoftVQ, and SoftCVQ, demonstrate notable improvements in protein structure reconstruction.

\subsection{Backbone Inpainting (Q2)}
\textit{How about applying protein language on generative task?}

\paragraph{Setting} We transform the whole CATH4.3 dataset as sequence of VQ-IDs using FoldTokenizer. We partition the dataset into training and testing sets, ensuring no overlap in CAT codes across proteins. During training, the length of the masked fragment in a protein of length $L$ is uniformly sampled from $U(5, L/3)$. The model is trained up to 20k epochs using the OneCycle scheduler and AdamW optimizer. The batch size is set to 128, the learning rate is 0.0005, and the padding length is 512.

\paragraph{Baselines} The sequence inpainting baselines are widely recognized ESM2 \citep{lin2022language} and the recent EvoDiff \citep{alamdari2023protein}. Structure inpainting methods are classified into angle-based and coordinate-based. Angle-based baselines include FoldingDiff \citep{wu2022protein} and DiffSDS \citep{gao2023diffsds}, while coordinate-based methods are ProtDiff \citep{trippe2022diffusion} , Chroma \citep{ingraham2023illuminating} and RFDiffusion \citep{watson2023novo}. FoldGPT belongs to the group of angle-based methods; in the future, we plan to extend it into the coordinate-based domain. Due to limited energy and computing resources, we utilize the pretrained checkpoints of baseline models. Consequently, the baselines may be overstated, given that the testing data may be part of their training set, and they \citep{lin2022language, alamdari2023protein, ingraham2023illuminating, watson2023novo} have utilized more training data.

\paragraph{Metrics} Denote ${\gS, \gX}$ and ${\hat{\gS}, \hat{\gX}}$ as the reference and predicted sequence-structure. We define $\hat{\gS} \rightarrow \hat{\gX}'$ as a folding process using ESMFold and $\hat{\gX} \rightarrow \hat{\gS}'$ inverse folding prosess using modified PiFold (considering only $C_{\alpha}$ atoms). Define $\texttt{Rec}(\gS_1, \gS_2)$ and $\texttt{TM}(\gX_1, \gX_2)$ as sequence recovery and structure TMScore. We report $\texttt{Rec}(\gS, \hat{\gS})$ and $\texttt{TM}(\gX, \hat{\gX}')$ for sequence inpainting. Similarly, we consider $\texttt{Rec}(\gS, \hat{\gS}')$ and $\texttt{TM}(\gX, \hat{\gX})$ for structure inpainting.

\textit{Refer to results on Table.\ref{tab:backbone_inpainting}, we provide following insights:}

\textbf{FoldGPT excels in sequence inpainting compared to baselines.} Despite using less training data than ESM2 and EvoDiff, FoldGPT achieves superior recovery and self-consistent TMScore. This highlights the effectiveness of the GPT-style generative model with learned protein language, outperforming MLM-style and diffusion-style approaches.

\textbf{FoldGPT outperforms other angle-based methods.} In structure inpainting, FoldGPT demonstrates significant improvement compared to FoldingDiff and DiffSDS. These results underscore the effectiveness of utilizing protein language for generative tasks with the same data representation.

\textbf{Coordinate-based representation is superior to angle-based. } Angle-based structure representation poses the risk of error accumulation, where atoms are added sequentially to the backbone using folding angles, making it challenging to consider pair-wise residue interactions. In light of this, we plan to extend FoldToken as a coordinate-based method.

\begin{figure}[h]
   \centering
   \includegraphics[width=3.3in]{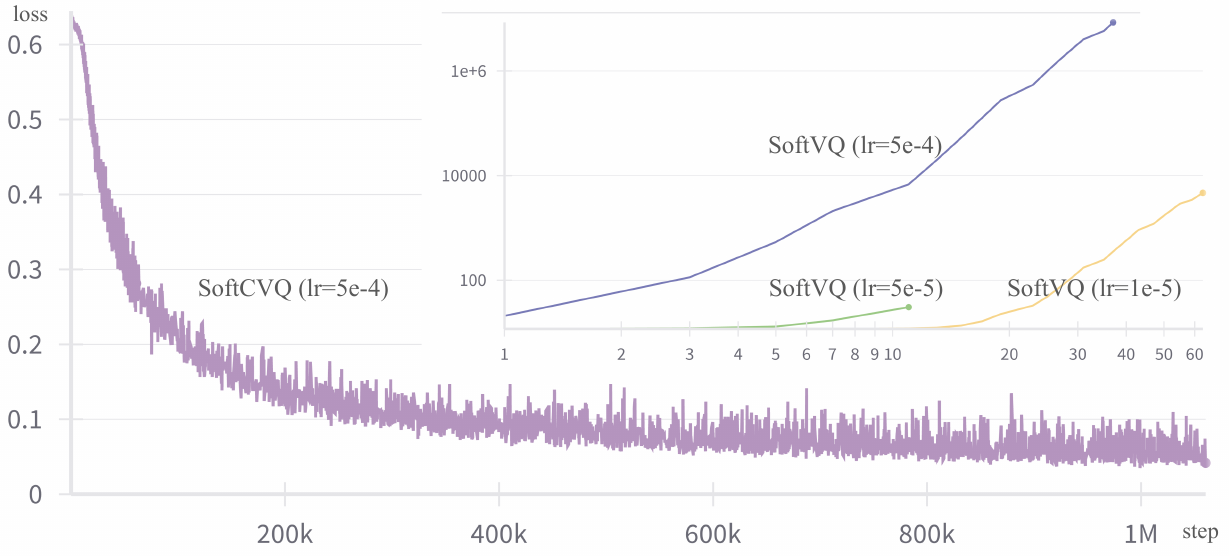}
   \caption{Binary VQ-ID is crucial for convergence. }
   \label{fig:inpaint_loss}
   \vspace{-5mm}
\end{figure}

\textbf{Binary VQ-ID is crucial for convergence.} As shown in Fig.\ref{fig:inpaint_loss}, attempting to use the language learned by SoftVQ with a class space of $2^{16}$ resulted in training crashes, regardless of learning rate adjustments. However, leveraging the binary VQ-ID learned by SoftCVQ, which decomposes the large class space into 16 binary subspaces and transforms the problem into 16 binary classifications, led to a successful and smooth convergence of the model.

\begin{table*}[t]
   \centering
   \resizebox{2 \columnwidth}{!}{\begin{tabular}{lcclcclcc}
      \toprule
              & \multicolumn{2}{c}{Sequence Design} &             & \multicolumn{2}{c}{Structure Design (angle-based)} &  & \multicolumn{2}{c}{Structure Design (coordinate-based)} \\
              & $\texttt{Rec}(\gS, \hat{\gS})$    & $\texttt{TM}(\gX, \hat{\gX}')$   &             & $\texttt{TM}(\gX, \hat{\gX})$   & $\texttt{Rec}(\gS, \hat{\gS}')$   &             & $\texttt{TM}(\gX, \hat{\gX})$   & $\texttt{Rec}(\gS, \hat{\gS}')$\\ \midrule
   ESM2    &   95.3     &       89.2                     & FoldingDiff    &   0.67       &      23.5            & ProtDiff    &   0.95       &      28.4               \\        
   EvoDiff    & 93.0       &    88.7            & DiffSDS & 0.74 &   25.6      & RFDiffusion & 0.99  &        33.9                         \\
   \chl FoldGPT & \chl 96.2      &  \chl    90.4          &\chl FoldGPT      &  \chl  0.80      &   \chl      28.9      & Chroma      &    0.98      &         34.5        \\ \bottomrule
   \end{tabular}}
   \caption{Results of backbone inpainting.}
   \label{tab:backbone_inpainting}
   \vspace{-3mm}
\end{table*}

\begin{table*}[b]
   \label{tab:kfold}
   \centering
   \resizebox{2 \columnwidth}{!}{
   \begin{tabular}{ccccccccc}
   \toprule
   \multirow{2}{*}{Method} & \multirow{2}{*}{Atoms} & \multirow{2}{*}{Antigen} & \multicolumn{2}{c}{CDR-H1} & \multicolumn{2}{c}{CDR-H2} & \multicolumn{2}{c}{CDR-H3} \\
   & & & AAR ($\uparrow$) & RMSD ($\downarrow$) & AAR ($\uparrow$) & RMSD($\downarrow$) & AAR ($\uparrow$) & RMSD($\downarrow$) \\
   \midrule
   LSTM & $\{C_{\alpha}, C, N\}$ & \XSolidBrush & 49.98$\pm$5.20\% & - & 28.50$\pm$1.55\% & - & 15.69$\pm$0.91\% & - \\ 
   RefineGNN        &  $\{C_{\alpha}, C, N\}$  & \XSolidBrush  & 39.40$\pm$5.56\%                           & 3.22$\pm$0.29       & 37.06$\pm$3.09\%                            & 3.64$\pm$0.40                           & 21.13$\pm$1.59\%                           & 6.00$\pm$0.55                           \\
   \chl FoldGPT &\chl $\{C_{\alpha}\}$ &\chl \XSolidBrush &\chl \textbf{50.75$\pm$5.91\%} &\chl \textbf{2.06$\pm$0.11} &\chl \textbf{39.71$\pm$3.02\%} &\chl \textbf{2.83$\pm$0.37} &\chl \textbf{23.37$\pm$2.92\%} &\chl \textbf{5.82$\pm$0.47}\\
   \midrule
   DiffAB       &  $\{C_{\alpha}, C, N, O\}$    &   \CheckmarkBold      & 61.34$\pm$1.98\%            &   1.02$\pm$0.66     & 37.66$\pm$1.89\%                      & 1.20$\pm$0.09            & 25.79$\pm$1.52\%                      & 3.02$\pm$0.11                         \\
   MEAN          &  $\{C_{\alpha}, C, N, O\}$   &  \CheckmarkBold     & 58.29$\pm$7.26\%                           & 0.98$\pm$0.16       & 47.15$\pm$3.09\%                           & 0.95$\pm$0.05                           & 36.38$\pm$3.08\%                           & 2.21$\pm$0.16                           \\
   ADesigner    &   $\{C_{\alpha}, C, N, O\}$   &     \CheckmarkBold        & \textbf{64.34}$\pm$3.37\% & \textbf{0.82}$\pm$0.12       & \textbf{55.52}$\pm$3.36\% & \textbf{0.79}$\pm$0.06 & \textbf{37.37}$\pm$2.33\% & \textbf{1.97}$\pm$0.19 \\
   \bottomrule
   \end{tabular}}
   \caption{Antibody design results. We report the mean (standard deviation) of 10-fold cross-validation for sequence and structure design.}
   \label{tab:antibody}
\end{table*}

\subsection{Application (Q3)}
\textit{How about applying protein language to antibody design?}
\paragraph{Setting \& Baselines} We fine-tuned FoldGPT on SAbDab to design CDR regions of antibodies. However, due to its angle-based structure representation, FoldGPT considered only the antibody and $C_{\alpha}$ atom, neglecting the antigen and other atoms. While previous studies have indicated that incorporating antibody context and additional backbone atoms could enhance performance, we defer this to the next version of FoldGPT, which will utilize a coordinate-based structure representation. Our antibody-only baselines include LSTM and RefineGNN. Additionally, we present antibody-antigen baselines such as DiffAB \citep{luo2022antigen}, MEAN \citep{kong2022conditional}, and ADesigner \citep{tan2023cross}.

\textit{Refer to the results on Table.\ref{tab:antibody}, we conclude that}

\textbf{FoldGPT shows promising performance in the antibody-only setting.} Despite baselines utilizing more backbone atoms, known to enhance recovery and RMSD \citep{tan2023cross, gao2023pifold}, FoldGPT consistently outperforms them in the tasks of CDR1, CDR2, and CDR3 design.

\textbf{While innovative, there is room for improvement in FoldGPT. } To our knowledge, FoldGPT stands as the first GPT-style model for sequence-structure co-generation. However, the current angle-based structure representation hinders the consideration of antibody-antigen complexes and the intricate 3D residue interactions, thereby limiting its potential to surpass state-of-the-art models. We plan to address these issues in the next version of FoldGPT.

\vspace{-3mm}
\section{Conclusion}
This paper presents a framework for learning a discrete protein language that describes sequence and structure simultaneously. We apply this learned language to generative tasks, specifically general backbone inpainting and antibody design, constructing the first GPT-style model for sequence-structure co-generation. Our results demonstrate promising outcomes compared to baselines under the same settings. We believe that our work could pave the way for new paradigms in protein representation and generation. Moreover, the proposed SoftCVQ is a general method with the potential to make broader impact in diverse domains, including CV, NLP, and graph learning.



\bibliography{icml24}
\bibliographystyle{icml2024}

\newpage
\appendix
\onecolumn
\section{Folding Angle Definition}
\label{appl:folding_angle}

\begin{figure}[h]
   \centering
   \includegraphics[width=3.5in]{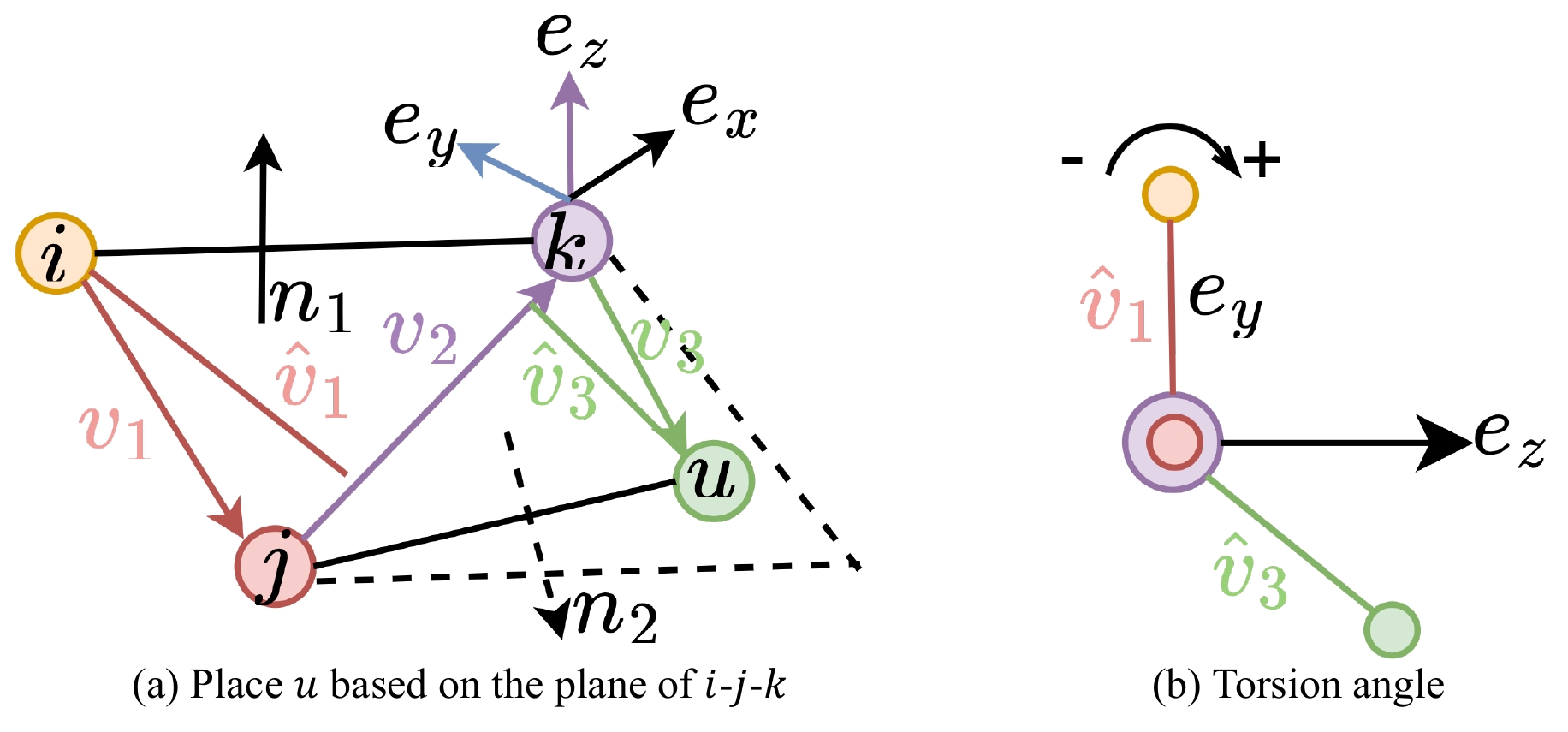}
   \caption{Torsion angle computation. }
   \label{fig:torsion_angle2}
\end{figure}

As shown in Fig.~\ref{fig:torsion_angle2}, following the backbone direction, given the previous three atoms, we can construct a local frame $T_{u} = \{\vx_i, \vx_j, \vx_k\}$. The next atom $\vx_u$ can be represented by the distance $r_u$ and two torsion angles $\alpha_u$ and $\beta_u$, denoted as $\va_u = (r_u, \alpha_u, \beta_u)$. 

\paragraph{Structure as angle sequence} The vanilla protein structures are represented as 3d points, where the rotation and translation equivariance should be considered in modeling. To directly apply the transformer-based encoder and decoder, we convert the 3d points into a sequence of angles, which is rotation and translation invariant. As shown in Fig.~\ref{fig:torsion_angle2}, following the backbone direction, given previous three atoms, we can construct a local frame $T_{u} = \{\vx_i, \vx_j, \vx_k\}$. The next atom $\vx_u$ can be represented by the distance $r_u$ and two torsion angles $\alpha_u$ and $\beta_u$, denoted as $\va_u = (r_u, \alpha_u, \beta_u)$. 
\begin{equation}
   \label{eq:torsion_angle}
   \begin{cases}
      \vv_1 = \vx_j - \vx_i \\
      \vv_2 = \vx_k - \vx_j \\
      \vv_3 = \vx_u - \vx_k \\
      \hat{\vv}_1 = -\vv_1 - ((-\vv_1) \cdot \vv_2) \vv_2 \\
      \hat{\vv}_3 = \vv_3 - (\vv_3 \cdot \vv_2) \vv_2 \\
      r_u = ||\vv_3||_2 \\
      \alpha_u = \angle (-\hat{\vv}_2, \hat{\vv}_3) \\
      \beta_u = \angle (\hat{\vv}_1, \hat{\vv}_3) = \angle (\vn_1, \vn_2)\\
   \end{cases}
\end{equation}

To completely represent the structure, we add virtual frames $T_0$ and $T_{n+1}$ at the start and end of the protein structure, where the virtual atoms have fixed $(r, \alpha, \beta)=(1,1,1)$ to its nearest backbone frames. Finally, the protein structure $\gX = \{\vx_i \in \sR^{3}: 1 \leq i \leq n\}$ will be represented as $\gS = \{(r_i, \alpha_i, \beta_i): 0 \leq i \leq n+1\}$.

\section{Local Metrics for Backbone Inpainting}
We define the average recovering loss per residue:
\begin{equation}
    \begin{cases}
       L_{r} = \frac{1}{b} \sum_{i=1}^b \sum_{j = 1}^{n_i} \frac{| r_{i,j}-\hat{r}_{i,j}|}{n_i}\\
       L_{\alpha} = \frac{1}{b} \sum_{i=1}^b \sum_{j = 1}^{n_i} \frac{| \alpha_{i,j}-\hat{\alpha}_{i,j}|}{n_i}\\
       L_{\beta} = \frac{1}{b} \sum_{i=1}^b \sum_{j = 1}^{n_i} \frac{| \beta_{i,j}-\hat{\beta}_{i,j}|}{n_i}\\
    \end{cases}
 \end{equation}
 where $b$ is batch size and $n_i$ is the length of protein $i$. The maximum recovering loss per residue is:
 \begin{equation}
    \begin{cases}
       \max L_{r} = \frac{1}{b} \sum_{i=1}^b \max_{j = 1}^{n_i} | r_{i,j}-\hat{r}_{i,j}|\\
       \max L_{\alpha} = \frac{1}{b} \sum_{i=1}^b \max_{j = 1}^{n_i} | \alpha_{i,j}-\hat{\alpha}_{i,j}|\\
       \max L_{\beta} = \frac{1}{b} \sum_{i=1}^b \max_{j = 1}^{n_i} | \beta_{i,j}-\hat{\beta}_{i,j}|\\
    \end{cases}
 \end{equation}

 \clearpage
 \section{Reconstruction Visualization}
\label{appl:visualization}
 \begin{figure}[h]
   \centering
   \includegraphics[width=6.5in]{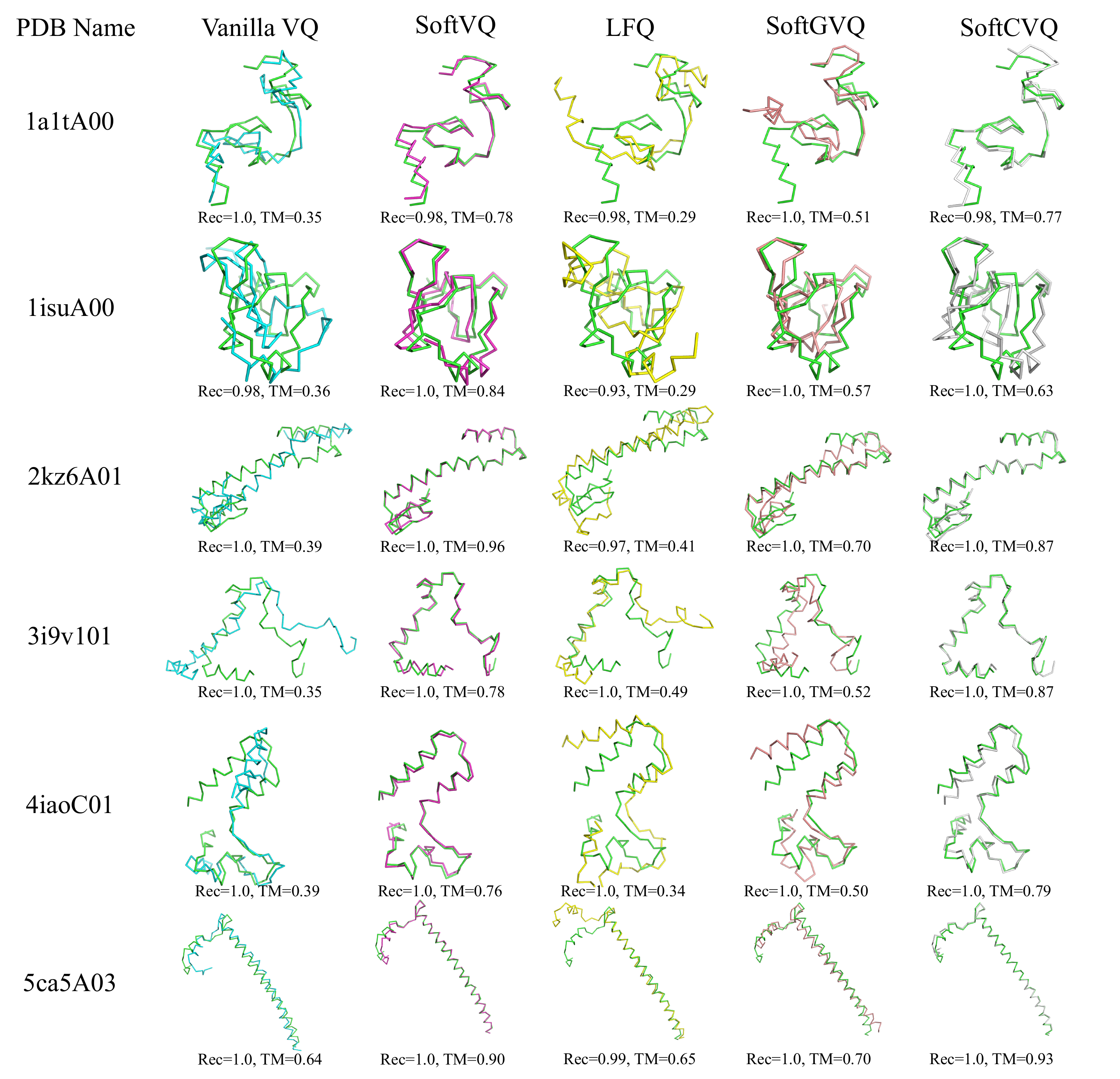}
   \caption{Reconstructed examples. }
   \label{fig:VQ_example}
\end{figure}

\clearpage
\section{Ablation of SoftCVQ}
\label{appl:ablation}
We investigate the impact of the number of MLP layers (\#MLP), the dimension of the codebook vector (\#CodeDim), and spherical normalization (sphere) on the reconstruction performance of SoftCVQ. Table \ref{tab:vq_ablation} demonstrates that appropriately increasing \#MLP and \#CodeDim enhances performance. Additionally, we highlight the significance of spherical normalization in achieving optimal results.
\begin{table}[h]
   \centering
   \resizebox{1.0 \columnwidth}{!}{
   \begin{tabular}{ccccccccccccccccccc}
   \toprule
   Method-ID & \multicolumn{3}{c}{Config} & \multicolumn{2}{c}{Global Metrics} & \multicolumn{2}{c}{Success Rates}    \\ 
    & \#MLP &\#CodeDim  &Sphere   & Rec    & TMScore & Rec $\geq 0.95$   & TMScore  $\geq 0.5$     \\ \midrule
   SCQ-1  &  2 & 16  &   & 0.9801 & 0.5853 & 0.8997 & 0.7638   \\
   SCQ-2  &  2 & 16  & \checkmark  & 0.9879 & 0.6451 & 0.9572  & 0.8892\\
   SCQ-3  &  2 & 32  &  & 0.9965 & 0.4285 & 0.9937 & 0.2184  \\
   SCQ-4  &  2 & 32  & \checkmark & 0.9896 & 0.6959 & 0.9561 & 0.9216 \\ 
   SCQ-5  &  3 & 32   &  & 0.9875 & 0.5867 & 0.9457 & 0.7367 \\ 
   SCQ-6  &  3 & 32   &  \checkmark & 0.9862 & 0.6697 & 0.9467 & 0.9028\\ 
   SCQ-7  &  6 & 32   &  \checkmark & 0.9880 & 0.7470 & 0.9603 & 0.9498\\ 
   SCQ-8  &  9 & 32   &  \checkmark & 0.9875 & 0.7015 & 0.9551 & 0.9101\\ 
   \bottomrule           
   \end{tabular}}
   \caption{Ablation of SoftCVQ on CATH4.3 test set.}
   \label{tab:vq_ablation}
\end{table}


\end{document}